\documentclass{osa-article}

\journal{osac}

\articletype{Research Article}

\begin{document}

\title{Topological edge states at singular points in non-Hermitian plasmonic systems}

\author{Yin Huang,\authormark{1,5} Yuecheng Shen,\authormark{2,6} and Georgios Veronis\authormark{3,4}}

\address{\authormark{1}Department of Optoelectrics Information Science and Engineering, School of Physics and Electronics, Central South University, Changsha, Hunan 410012, China\\
\authormark{2}State Key Laboratory of Optoelectronic Materials and Technologies, School of Electronics and Information Technology, Sun Yat-Sen University, Guangzhou, Guangdong 510275, China}
\address{\authormark{3}School of Electrical Engineering and Computer Science, Louisiana State University, Baton Rouge, LA 70803, USA\\
\authormark{4}Center for Computation and Technology, Louisiana State University, Baton Rouge, LA 70803, USA\\
\authormark{5}yhuan15@csu.edu.cn \\
\authormark{6}shenyuecheng@mail.sysu.edu.cn}




\begin{abstract}
We introduce non-Hermitian plasmonic waveguide-cavity systems with topological edge states (TESs) at singular points. The compound unit cells of the structures consist of metal-dielectric-metal (MDM) stub resonators side-coupled to an MDM waveguide. We show that we can realize both a TES and an exceptional point at the same frequency when a proper amount of loss is introduced into a finite three-unit-cell structure. We also show that the finite structure can exhibit both a TES and a spectral singularity when a proper amount of gain is introduced into the structure. In addition, we show that we can simultaneously realize a unidirectional spectral singularity and a TES when proper amounts of loss and gain are introduced into the structure. We finally show that this singularity leads to extremely high sensitivity of the reflected light intensity to variations of the refractive index of the active materials in the structure. TESs at singular points could potentially contribute to the development of singularity-based plasmonic devices with enhanced performance.
\end{abstract}

\section{Introduction}
Exceptional points (EPs) are singular points in the spectra of non-Hermitian Hamiltonians in open quantum systems associated with the coalescence of the eigenvalues of the system and their corresponding eigenvectors \cite{Moiseyev,Heiss:04}. Unidirectional reflectionless light propagation in non-Hermitian optical systems reveals the presence of EPs \cite{Lin:11,Feng:13,Ge:12,Huang:15,Yin:13,Huang:17}. Another type of singular points are spectral singularities (SSs) associated with the lack of completeness of the eigenvectors of non-Hermitian Hamiltonians in the continuous spectra \cite{Mostafazadeh:09, Konotop:19}. The reflection of non-Hermitian optical systems at SSs tends to infinity and corresponds to lasing at threshold gain \cite{Mostafazadeh:11}. In addition, Ramezani $et$ $al$ introduced another type of singular points known as unidirectional spectral singularities at which the system exhibits zero reflection from one side and infinite reflection from the opposite side \cite{Ramezani:14}. Singular points can lead to novel optical devices such as optical network analyzers \cite{Feng:13}, switches \cite{Doppler:16, Zhao:16}, directional lasers \cite{Peng:16}, isolators \cite{Laha:20}, perfect absorbers \cite{Huang:16}, slow-light waveguides \cite{Goldzak:18}, and sensors \cite{Chen:17,Hodaei:17,Huang:19,Sakhdari:17,Farhat:20} with applications in optical computation, communication, and information processing.

Topological insulators are peculiar electronic materials that possess non-trivial topological states on their edge or surface \cite{Bernevig:06, Hasan:10}. Topological edge states (TESs) are insensitive to disorder and can lead to field intensity enhancement \cite{Bernevig:06, Hasan:10}. There have been both theoretical and experimental demonstrations that topological concepts can be transferred to photonics \cite{Haldane:08, Wang:09, Hafezi:11, Fang:12, Lang:12}. Recently, Poshakinskiy $et$ $al$ demonstrated that TESs can survive in non-Hermitian systems consisting of one-dimensional resonant photonic crystals with modulated couplings, and presented a method to detect these edge states \cite{Poshakinskiy:14}. This opens a pathway to study the connection between TESs and EPs. Most recently, Zhu $et$ $al$ simultaneously observed TESs and EPs in non-Hermitian acoustic systems with judiciously tailored losses \cite{Zhu:18}. It would be interesting to investigate whether it is possible to realize both a TES and an EP at the same frequency in non-Hermitian optical systems. In addition, to date the connection between edge states and other types of singular points, such as spectral singularities, has not been explored. It would also be interesting to consider simultaneously realizing edge states and other types of singular points in non-Hermitian optical systems. This could further boost the potential of singular-point-based devices in photonic applications.

In this paper, we introduce non-Hermitian plasmonic waveguide-cavity systems with TESs at singular points. We first calculate the eigenfrequencies of TESs in a periodic plasmonic structure based on the Aubry-Andre-Harper (AAH) model with compound unit cells consisting of MDM stub resonators side-coupled to an MDM waveguide. The AAH model is the one-dimensional momentum-space projection of the integer quantum Hall effect, and therefore exhibits nontrivial topological properties \cite{Lang:12, Kraus:12, Liu:15}. We show that we can realize both a TES and an EP at the same frequency when a proper amount of loss is introduced into a finite plasmonic structure consisting of three compound unit cells. We also show that the finite structure can exhibit both a TES and a SS at the same frequency when a proper amount of gain is introduced into the structure. In addition, we show that we can simultaneously realize a unidirectional spectral singularity and a TES when proper amounts of loss and gain are introduced into the finite plasmonic structure. We finally show that this singularity can lead to extremely high sensitivity of the reflected light intensity to variations of the refractive index of the active materials in the structure.

\section{Model}
\begin{figure}[htb]
\centering\includegraphics[width=9cm]{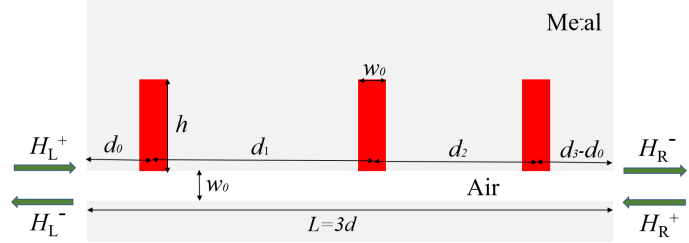}
\caption{Schematic of the compound unit cell of a periodic structure consisting of an MDM waveguide side coupled to three identical MDM stub resonators.}
\end{figure}
Analogous to the AAH model, we consider a periodic plasmonic structure with compound unit cells consisting of metal-dielectric-metal (MDM) stub resonators side-coupled to an MDM waveguide (Fig. 1). Among the different plasmonic waveguiding geometries, MDM plasmonic waveguides have attracted a lot of interest \cite{Yu:08} because they support modes with deep subwavelength size over a very broad range of frequencies extending from DC to visible \cite{Economou:69}, and are relatively easy to fabricate \cite{Salamin:15}. The compound unit cells consist of $N$ side-coupled MDM stub resonators. The distance between the $n$th and ($n+1$)th side-coupled stubs in the compound unit cell is modulated and given by
\begin{equation}
d_n=d\{1+\eta \cos[2\pi b(n-1)+\phi]\}.
\end{equation}
Here, $b=1/N$, $d$ is the distance without modulation, and $\eta$ is the
modulation strength. Figure 1 shows a compound unit cell with three side-coupled stubs, so that $b=1/3$ ($N=3$). In addition, $\phi$ is an arbitrary phase. When $\phi$ spans the interval between $-\pi$ and $\pi$, one-dimensional problems can be mapped to two-dimensional integer quantum Hall effect problems with the Landau gauge characterized by nonzero Chern numbers \cite{Lang:12}.

The periodic system of Fig. 1 can be described by temporal coupled mode theory (CMT) for the mode amplitudes $a_n$ of the side-coupled stubs, as follows
\begin{equation}
j\omega a_n=(j\omega_0-\frac{1}{\tau_0}-\frac{1}{\tau})a_n-\frac{1}{\tau_0} \sum_{n^{\prime} \neq n}\Lambda_{n n^{\prime}}a_n,
\end{equation}
where $\omega_0$ is the resonant frequency, $\frac{1}{\tau_0}$ is the decay rate of the stub resonator field due to the power escape through the waveguide, $\frac{1}{\tau}$ is the decay (growth) rate due to the internal loss (gain) in the stub resonator, $\Lambda_{n n^{\prime}}=e^{-\gamma|x_n-x_{n^\prime}|}$ with the $n$th side-coupled stub centered at $x_n$, and $\gamma$ is the complex propagation constant of the propagating fundamental mode of the MDM waveguide.

The properties of the compound unit cell of Fig. 1 can be described by the transfer matrix $\mathbf{M}$ defined by the following equation
\begin{equation}
\left[
\begin{array}{c}
H_R^-\\
H_R^+\\
\end{array}
\right]
=
\mathbf{M}\left[
\begin{array}{c}
H_L^+\\
H_L^-\\
\end{array}
\right]=\left[
\begin{array}{c c}
M_{11} & M_{12} \\
M_{21} & M_{22} \\
\end{array}
\right]\left[
\begin{array}{c}
H_L^+\\
H_L^-\\
\end{array}
\right].
\end{equation}
where $H^+_L$ , and $H^-_L$ are the complex magnetic field amplitudes of the incoming and outgoing modes at the left port, respectively. Similarly, $H^+_R$ , and $H^-_R$ are the complex magnetic field amplitudes
of the incoming and outgoing modes at the right port, respectively (Fig. 1). The transfer matrix $\mathbf{M}$ can be calculated by
\begin{equation}
\mathbf{M}=\mathbf{M}_4\mathbf{M}_s\mathbf{M}_3\mathbf{M}_s\mathbf{M}_2\mathbf{M}_s\mathbf{M}_1,
\end{equation}
where
$\mathbf{M}_s=
\left[
\begin{array}{c c}
t_s-\frac{r_s^2}{t_s} & \frac{r_s}{t_s} \\
-\frac{r_s}{t_s} & \frac{1}{t_s} \\
\end{array}
\right]$ is the transfer matrix of a system consisting of an MDM waveguide side-coupled to a stub, while $r_s=-\frac{1/\tau_0}{j(\omega-\omega_0)+1/\tau+1/\tau_0}$ and $t_s=1+r_s$ are the complex reflection and transmission coefficients of the system, respectively \cite{Joannopoulos}.
In addition, $\mathbf{M}_i=\left[
\begin{array}{c c}
e^{-\gamma L_i} & 0 \\
0 & e^{\gamma L_i} \\
\end{array}
\right], i=1, 2, 3, 4$, where $L_1=d_0$, $L_2=d_1$, $L_3=d_2$, $L_4=d_3-d_0$, and $d_0$ is an arbitrary distance with $d_0<d_3$. The transmission coefficient for the compound unit cell is given by
\begin{equation}
t_1=\frac{H_R^-}{H_L^+}\big|_{H_R^+=0}=\frac{H_L^-}{H_R^+}\big|_{H_L^+=0}=M_{11}-\frac{M_{12}M_{21}}{M_{22}},
\end{equation}
while the left and right reflection coefficients for the compound unit cell can be obtained by
\begin{equation}
r_{1l}=\frac{H_L^-}{H_L^+}\big|_{H_R^+=0}=-\frac{M_{21}}{M_{22}}, r_{1r}=\frac{H_R^-}{H_R^+}\big|_{H_L^+=0}=\frac{M_{12}}{M_{22}}.
\end{equation}
In addition, the dispersion relation of the periodic plasmonic structure with the compound unit cell of Fig. 1 can be computed using the transfer matrix $\mathbf{M}$ of the unit cell and the Bloch boundary conditions $H^-_R=e^{jkL}H^+_L$, $H^+_R=e^{jkL}H^-_L$ by
\begin{equation}
\bigg|\mathbf{M}- \left[
\begin{array}{c c}
e^{jkL} & 0 \\
0 & e^{jkL} \\
\end{array}
\right]\bigg|=0,
\end{equation}
where $L=3d$ is the overall length of the compound unit cell (Fig. 1) and $k$ is the Bloch wave vector.

\section{Results}
\subsection{Topological edge states}
\begin{figure}[htb]
\centering\includegraphics[width=11cm]{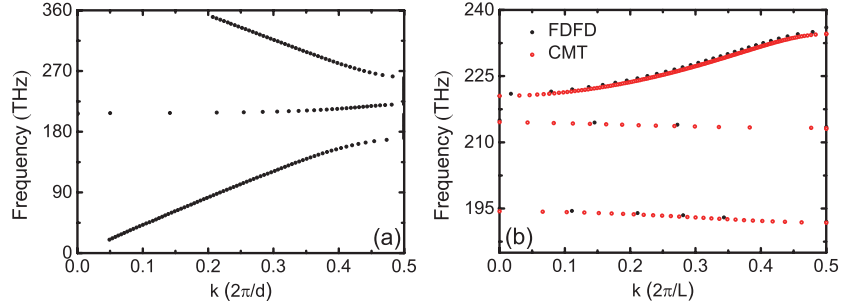}
\caption{(a) Dispersion relation of the periodic structure without modulation calculated using FDFD. Results are shown for $d_1=d_2=d_3=d=500$ nm, $h=60$ nm, $w_0=50$ nm (Fig. 1), and $\eta=0$ [Eq. (1)]. The stubs are filled with InGaAsP with InAs quantum dots, and the metal is silver. Here, we assume that silver is lossless. (b) Dispersion relation of a periodic structure with the compound unit cell of Fig. 1, in which the distances between adjacent side-coupled stubs are modulated as in Eq. (1), calculated using FDFD (black circles) and CMT (red circles). Results are shown for $d_1=d(1+\eta \cos\phi)$, $d_2=d[1+\eta \cos(\frac{2\pi}{3}+\phi)]$, $d_3=d[1+\eta \cos(\frac{4\pi}{3}+\phi)]$, $\eta=0.4$, and $\phi=0$. All other parameters are as in Fig. 2(a).}
\end{figure}

We first consider a periodic plasmonic structure without modulation [$\eta=0$ in Eq. (1) so that $d_1=d_2=d_3=d$]. We choose $d=500$ nm, $h=60$ nm, and $w_0=50$ nm (Fig. 1). The metal is silver and the stubs are filled with InGaAsP with InAs quantum dots. In this subsection, we assume that silver is lossless. Please note, however, that the material loss in silver is included in Subsections 3.2, 3.3, and 3.4. In the presence of pumping, InGaAsP with InAs quantum dots exhibits optical gain \cite{Yu:08, Babicheva:12}.
The real part of the dielectric constant of InGaAsP is 11.38 \cite{Babicheva:12}. In Fig. 2(a), we show the dispersion relation of the structure without modulation calculated using the finite-difference frequency-domain (FDFD) method. The structure supports a middle band corresponding to a mode with slow group velocity. We also consider a structure in which the distances between adjacent side-coupled stubs are modulated as in Eq. (1) with $b=1/3$, $\eta=0.4$, and $\phi=0$. The compound unit cell has overall length $L=3d$ (Fig. 1). The middle band with slow group velocity of the structure without modulation [Fig. 2(a)] splits into three bands due to band mixing for the structure with modulation [Fig. 2(b)]. Figure 2(b) also shows the dispersion relation of the structure with modulation calculated using CMT [Eq. (7)] (red circles). We observe that there is very good agreement between the CMT results and the exact results obtained using FDFD.

\begin{figure}[htb]
\centering\includegraphics[width=8cm]{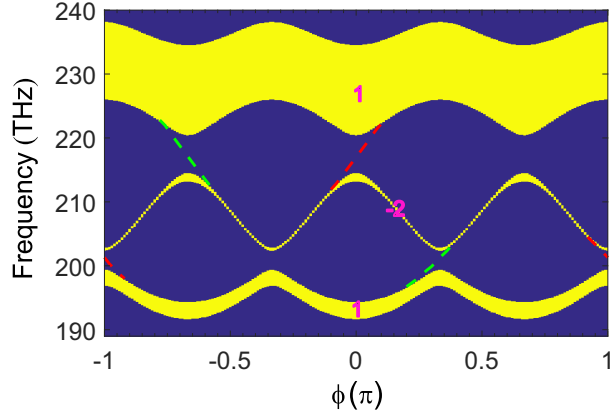}
\caption{Projected dispersion relation of the periodic structure with the compound unit cell of Fig. 1 as a function of $\phi$ calculated with CMT. The yellow regions correspond to the bands separated by the band gaps (blue regions). Also shown is the Chern number of each band. The eigenfrequencies of the edge states for right (left) incidence are shown with red (green) dashed line. All other parameters are as in Fig. 2(b).}
\end{figure}

Figure 3 shows the projected band structure of the periodic structure with the compound unit cell of Fig. 1 for $\eta=0.4$ and $b=1/3$ as a function of the phase $\phi$ [Eq. (1)], calculated with CMT.
For a fixed $\phi$, we have three bands (yellow regions) in the frequency range of interest. As an example, for $\phi=0$ the system supports three bands with frequencies ranging from 192 THz to 195 THz, 214 THz to 215 THz, and 221 THz to 235 THz [Figs. 2(b) and 3]. As the phase $\phi$ varies from $-\pi$ to $\pi$, the Bloch eigenstates of the system are a function of $k$ and $\phi$. Thus, each band is characterized by a Chern number which is defined as $\frac{1}{2\pi j}\int_{-\pi}^\pi d\phi \int_{-\pi/L}^{\pi/L} dk(\partial_kA_\phi-\partial_\phi A_k)$, where $A_k=\sum_s a_s^*\partial_k a_s$, $A_\phi=\sum_s a_s^*\partial_\phi a_s$, and $a_s, s=1, 2, 3$, are the mode amplitudes satisfying the Bloch
condition $a_{s+3l}(k,\phi)=e^{jlkL}a_{s}(k,\phi)$ for $l=0,\pm 1, \pm 2,\ldots$. This Chern number can also be deduced from the winding number of the band gap using the phase of the reflection coefficient for the semi-infinite structure \cite{Zhu:18,Poshakinskiy:15}. Imposing the Bloch boundary condition, the reflection coefficient for the semi-infinite structure $r_\infty$, when the waveguide mode is incident from the right, can be obtained using the transfer matrix
\begin{equation}
r_\infty=\frac{e^{-jkL}-M_{22}}{M_{21}}.
\end{equation}
Note that, in the lossless case, the left and right reflection coefficients are equal.

Figure 4(a) shows the absolute value of the reflection coefficient $|r_\infty|$ calculated using Eq. (9), when the waveguide mode is incident from the right onto the semi-infinite plasmonic structure consisting of compound unit cells as in Fig. 1. We observe that $|r_\infty|$ shown in Fig. 4(a) is consistent with the dispersion relation shown in Fig. 3. The absolute value of the reflection coefficient $|r_\infty|$ for frequencies lying inside the band gaps is 1 [Fig. 4(a)]. Hence, the reflection coefficient $r_\infty$ for frequencies lying inside the band gaps is $1e^{j\theta}$, where $\theta$ is the phase of the reflection coefficient. The winding number, which is the topological invariant of the band gap, can thus be calculated using \cite{Asboth}
\begin{equation}
w=\frac{1}{2\pi j}\int_0^{2\pi}\frac{\partial \ln[r_\infty(\phi)]}{\partial \phi}d\phi=\frac{1}{2\pi }\int_0^{2\pi}\partial\theta(\phi).
\end{equation}

Figure 4(b) shows the phase $\theta$ of the reflection coefficient $r_\infty$ when the waveguide mode is incident from the right onto the semi-infinite structure. The extra phase that the reflection coefficient accumulates when $\phi$ varies from $-\pi$ to $\pi$ is 0, $2\pi$, $-2\pi$, and 0 for the first, second, third, and fourth band gap, respectively, in the frequency range of interest. The winding numbers of these four band gaps are therefore 0, 1, -1, and 0 [Eq. (8)]. Thus, since the Chern number of a band is equal to the winding number of the above-lying band gap minus the winding number of below-lying band gap \cite{Poshakinskiy:15}, the Chern numbers of the three bands in the frequency range of interest are 1, -2, and 1 (Fig. 3). Non-zero Chern numbers indicate the existence of TESs in the band gaps based on the bulk-boundary correspondence \cite{Asboth}.

\begin{figure}[htb]
\centering\includegraphics[width=12cm]{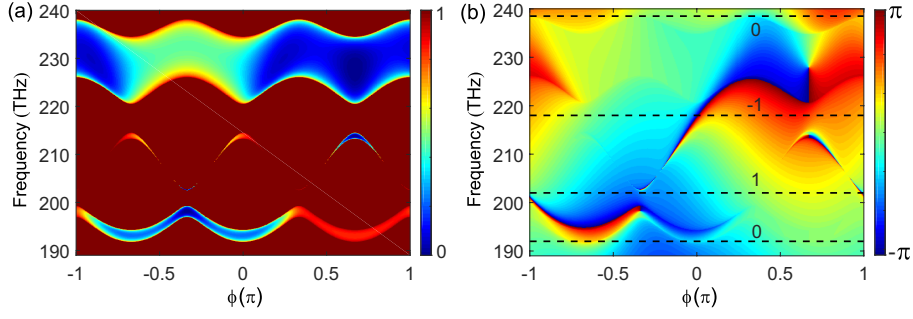}
\caption{(a) The absolute value of the reflection coefficient $|r_\infty|$ as a function of $\phi$, when the waveguide mode is incident from the right onto the semi-infinite plasmonic structure consisting of compound unit cells as in Fig. 1, calculated using CMT. All parameters are as in Fig. 3. (b) The phase $\theta$ of the reflection coefficient $r_\infty$ as a function of $\phi$, when the waveguide mode is incident from the right onto the semi-infinite structure. Also shown is the winding number of each band gap. All parameters are as in Fig. 3.}
\end{figure}

The TESs can be obtained by direct calculation of Eq. (2). However, this calculation is challenging since $\Lambda_{n n^{\prime}}$ depends on the frequency $\omega$. Alternatively, the TESs can be extracted from the zeros of the reflection coefficient $r_\infty$ when an appropriate amount of loss ($\frac{1}{\tau}>0$) is introduced into the semi-infinite structure; that is, $r_{\infty}(\omega, \frac{1}{\tau})=0$ \cite{Poshakinskiy:14,Zhu:18}. The underlying physical mechanism behind this can be explained as follows: when light with the eigenfrequency of the edge state is incident on the structure, the edge state is excited and the light is deeply trapped in the edge region. If an appropriate amount of loss is introduced into the system, the trapped light is completely absorbed, and thus $r_{\infty}(\omega, \frac{1}{\tau})=0$. The zeros of $r_{\infty}(\omega, \frac{1}{\tau})$ are identical to the zeros of $r_1(\omega, \frac{1}{\tau})$ for a unit cell \cite{Zhu:18}. Thus, the eigenfrequencies of the left and right edge states can be retrieved from $r_{1l}(\omega, \frac{1}{\tau})=0$ and $r_{1r}(\omega, \frac{1}{\tau})=0$, respectively. Figure 3 shows that the structure possesses two chiral edge states in each band gap. The green dashed line corresponds to the left edge state [$r_{1l}(\omega, \frac{1}{\tau})=0$], while the red dashed line corresponds to the right edge state [$r_{1r}(\omega, \frac{1}{\tau})=0$]. The chirality of the edge states is a property of two-dimensional integer quantum Hall systems. AAH systems are the one-dimensional momentum-space projection of two-dimensional integer quantum Hall systems and exhibit similar topological properties \cite{Lang:12}.

\subsection{Topological edge states at exceptional points}

\begin{figure}[htb]
\centering\includegraphics[width=11.5cm]{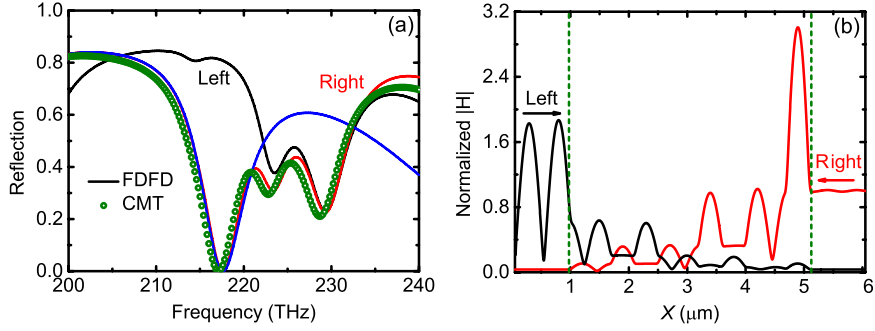}
\caption{(a) Reflection spectra when the waveguide mode is incident from the right onto the single-unit-cell structure calculated using FDFD (blue solid line). Also shown are the reflection spectra when the waveguide mode is incident from the left (black solid line) and right (red solid line) onto the three-unit-cell structure calculated using FDFD. In addition, shown are the calculated reflection spectra using CMT, when the waveguide mode is incident from the right onto the three-unit-cell structure (green circles). Here, the material loss in silver is included. The dielectric constant of the material filling the stubs (InGaAsP with InAs quantum dots) is equal to $11.38+j0.166$. All other parameters are as in Fig. 2(b). (b) Profile of the magnetic field amplitude in the middle of the MDM waveguide, normalized with respect to the field amplitude of the incident waveguide mode in the middle of the waveguide, when the mode is incident from the left (black solid line) and right (red solid line) onto the three-unit-cell structure at $f=217.5$ THz. All other parameters are as in Fig. 5(a). The vertical dashed lines indicate the boundaries between the three-unit-cell structure and the MDM waveguide.}
\end{figure}

In this subsection, to realize both a TES and an EP at the same frequency, we consider a finite structure consisting of three compound unit cells as in Fig. 1 with $\phi=0$ [Eq. (1)]. We found that three unit cells are sufficient for the finite structure to exhibit characteristics of the infinite periodic structure \cite{Zhu:18}. As shown in Fig. 3, a right TES exists in the second band gap around $\phi=0$ (red dashed line), when a proper amount of loss is introduced into the system. More specifically, when the waveguide mode is incident onto a single-unit-cell structure from the right, we find using CMT that $r_{1r}=0$ for decay rate due to internal loss in the stub resonators $1/\tau$ equal to $8.77\times10^{12} {\rm sec}^{-1}$. We also find that, when we take into account the material loss of silver, the material filling the stubs (InGaAsP with InAs quantum dots) must exhibit gain with the imaginary part of its dielectric constant equal to 0.166, in order for $1/\tau$ to be equal to $8.77\times10^{12} {\rm sec}^{-1}$. Figure 5(a) confirms that the reflection, when the waveguide mode is incident from the right onto the single-unit-cell structure, calculated using FDFD (blue solid line) becomes zero ($r_{1r}=0$) for the dielectric constant of the material filling the stubs equal to $11.38+j0.166$ at the frequency of $f=217.5$ THz. We also find that the calculated reflection using FDFD, when the waveguide mode is incident from the right onto the three-unit-cell structure (red solid line), becomes zero ($r_{3r}=0$) at the same frequency ($f=217.5$ THz) as the single-unit-cell structure [Fig. 5(a)]. Thus, we conclude that there exists a TES localized at the right boundary of the three-unit-cell structure for $f=217.5$ THz. Figure 5(a) also shows the calculated reflection using CMT, when the waveguide mode is incident from the right onto the three-unit-cell structure (green circles). Once again, we observe that there is very good agreement between the CMT results and the exact results obtained using FDFD.

In addition, we find that for the three-unit-cell structure the reflection for the waveguide mode incident from the right (red solid line) is zero at $f=217.5$ THz, while the reflection for the waveguide mode incident from the left (black solid line) is nonzero [Fig. 5(a)]. Thus, the structure exhibits unidirectional reflectionless propagation at $f=217.5$ THz. The properties of the three-unit-cell structure can also be described by the scattering matrix $
\mathbf{S}=\left[
\begin{array}{c c}
t_3 & r_{3r} \\
r_{3l} & t_3 \\
\end{array}
\right]$. The matrix $S$ is non-Hermitian in the presence of loss, and its complex eigenvalues are $\lambda_s^\pm=t_3\pm\sqrt{r_{3r}r_{3l}}$. The corresponding eigenstates, which are $\psi_\pm=(1,\pm\sqrt{\frac{r_{3r}}{r_{3l}}})$ for $r_{3l}\neq0$, are not orthogonal \cite{Huang:17}. In the case of unidirectional reflectionless propagation in the right direction ($r_{3r}=0, r_{3l}\neq 0$), both the scattering matrix $\mathbf{S}$ eigenvalues and their corresponding eigenstates coalesce, and thus an EP is formed. In other words, the three-unit-cell structure exhibits both a TES and an EP at the same frequency. In addition, we can observe both the localized TES and the unidirectional reflectionless propagation in the normalized magnetic field distributions at that frequency. When the waveguide mode is incident from the left, there is strong reflection, so that the incident and reflected modal fields form a strong interference pattern [Fig. 5(b)]. On the other hand, when the waveguide mode is incident from the right, there is hardly any reflection. In addition, the field is enhanced at the right edge, demonstrating the existence of the edge state [Fig. 5(b)].

\subsection{Topological edge states at spectral singularities}
\begin{figure}[htb]
\centering\includegraphics[width=11.5cm]{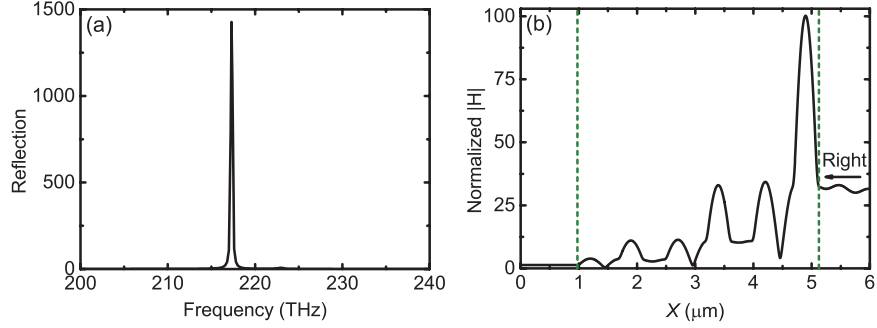}
\caption{(a) Reflection spectra when the waveguide mode is incident from the right onto the three-unit-cell structure calculated using FDFD. The material loss in silver is included. The dielectric constant of the material filling the stubs (InGaAsP with InAs quantum dots) is equal to $11.38+j0.58$. All other parameters are as in Fig. 2(b). (b) Profile of the magnetic field amplitude in the middle of the MDM waveguide, normalized with respect to the field amplitude of the incident waveguide mode in the middle of the waveguide, when the mode is incident from the right onto the three-unit-cell structure at $f=217.5$ THz. All other parameters are as in Fig. 6(a). The vertical dashed lines indicate the boundaries between the three-unit-cell structure and the MDM waveguide.}
\end{figure}
Another type of singular points are spectral singularities which, unlike EPs, can only occur in the presence of gain \cite{Mostafazadeh:09}. The reflection of non-Hermitian optical systems at SSs tends to infinity \cite{Mostafazadeh:11}. Based on Eqs. (3), (4) and (8), the reflection coefficient for the semi-infinite structure $r_\infty$ depends on the reflection coefficient for the single stub structure $r_s$ which is a function of $\omega-j\frac{1}{\tau}$. The winding number of a band gap can thus be written as a function of $\omega-j\frac{1}{\tau}$
\begin{equation}
w(\omega-j\frac{1}{\tau})=\frac{1}{2\pi j}\int_0^{2\pi}\frac{\partial \ln[r_\infty(\omega-j\frac{1}{\tau}, \phi)]}{\partial \phi}d\phi.
\end{equation}
In the lossless case ($\frac{1}{\tau}$=0), the winding number of the third band gap is $-1$ [Fig. 4(b)]. On the other hand, for $\frac{1}{\tau}\rightarrow\infty$, $r_s=-\frac{1/\tau_0}{j(\omega-\omega_0)+1/\tau+1/\tau_0}\rightarrow 0$, which indicates that the waveguide-cavity structure reduces to a straight waveguide. The reflection coefficient $r_\infty$ therefore becomes zero and is independent of $\phi$. Thus, based on Eq. (10), $\lim_{\frac{1}{\tau}\rightarrow\infty}w=0$. Since the winding number as a topological invariant is an integer, its value must abruptly change from $-1$ to $0$ at certain finite $\frac{1}{\tau} \in (0,\infty)$. Such a discontinuity can only be caused by a singularity of the integrand in Eq. (10), which corresponds to a zero of the reflection coefficient $r_\infty$, that is, $r_\infty(\omega-j\frac{1}{\tau})=0$ \cite{Poshakinskiy:15}. In the lossless case ($\frac{1}{\tau}$=0), $|r_\infty|^2=1$ for frequencies lying inside the band gap. In the presence of loss, $|r_\infty|^2=1$ turns into $r_\infty(\omega-j\frac{1}{\tau})r_\infty^*(\omega+j\frac{1}{\tau})=1$. Since $r_\infty(\omega-j\frac{1}{\tau})=0$, we must have $r_\infty(\omega+j\frac{1}{\tau})\rightarrow\infty$. In other words, if the reflection in the semi-infinite structure tends to infinity when a proper amount of gain is introduced, the structure exhibits a TES. In addition, such a pole of the reflection coefficient $r_\infty$ corresponds to the presence of a SS \cite{Mostafazadeh:11}. Thus, this analysis suggests that, if a proper amount of gain is introduced into the structure, it can exhibit both a TES and a SS at the same frequency.

To implement this in plasmonic waveguide-cavity systems as in Fig. 1, we consider as before a finite structure consisting of three compound unit cells with $\phi=0$ for $d=500$ nm, $h=60$ nm, $w_0=50$ nm, and $\eta=0.4$ [Eq. (1)]. Using CMT, we find that the reflection coefficient for the waveguide mode incident onto the finite structure from the right $|r_{3r}|$ tends to infinity, when the decay rate of the stub resonator mode amplitude $\frac{1}{\tau}$ approximately equals to $-1.37\times10^{13} {\rm sec}^{-1}$. To satisfy this condition we find that, when we take into account the material loss of silver, the material filling the stubs (InGaAsP with InAs quantum dots) must exhibit gain with the imaginary part of its dielectric constant equal to 0.58. Figure 6(a) shows that the calculated reflection using FDFD when the waveguide mode is incident from the right onto the three-unit-cell structure becomes extremely large for the dielectric constant of the material filling the stubs equal to $11.38+j0.58$ at $f=217.5$ THz. This result indicates that a TES exists at the right boundary of the structure for $f=217.5$ THz. In addition, the narrow-width resonance corresponding to a lasing process suggests that the structure exhibits a SS. The TES and right lasing can also be observed in the normalized magnetic field distributions at the SS [Fig. 6(b)]. When the waveguide mode is incident from the right, the reflected fields are greatly amplified. In addition, the greatly enhanced field at the right edge of the structure confirms the existence of the edge state.

\subsection{Topological edge states at unidirectional spectral singularities}
A unidirectional spectral singularity is a singular point at which zero reflection from one side and infinite reflection from the opposite side are simultaneously realized \cite{Ramezani:14}. As we saw in Subsection 3.3, the edge state at the SS shows up as a narrow-width resonance in the reflection spectra when a proper amount of gain is introduced into the stub resonators. Based on Eqs. (6) and (8), it is possible that the reflection coefficient in the right direction for the semi-infinite structure $r_\infty$ approaches infinity, and the reflection coefficient in the left direction for the structure consisting of a single unit cell is zero. This suggests that a properly designed structure could exhibit both a TES and a unidirectional spectral singularity at the same frequency.

\begin{figure}[htb]
\centering\includegraphics[width=10.5cm]{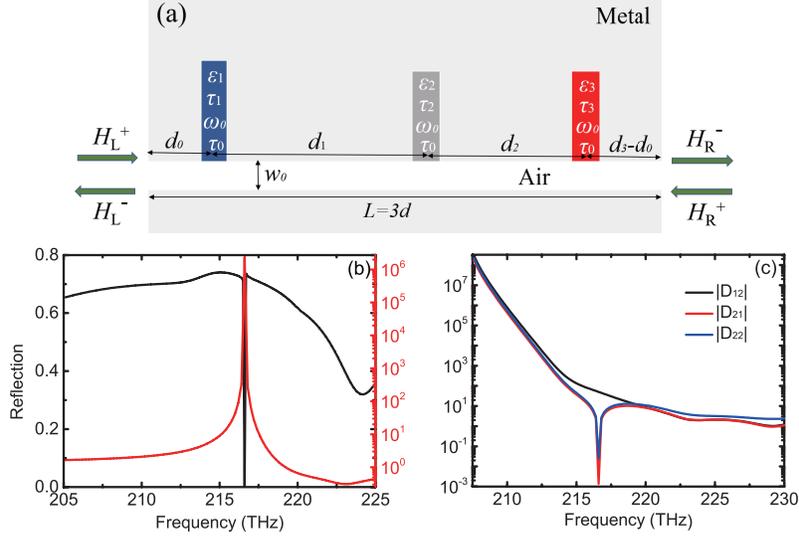}
\caption{(a) Schematic of the compound unit cell of a structure consisting of a finite number of unit cells. The unit cell consists of an MDM waveguide side coupled to three MDM stub resonators which have the same resonance frequency $\omega_0$ and the same decay rate due to power escape through the waveguide $\frac{1}{\tau_0}$, but different decay (growth) rates due to the internal loss (gain) in the stub resonators $\frac{1}{\tau_i}, i=1, 2, 3$. (b) Reflection spectra when the waveguide mode is incident from the left (black solid line) and right (red solid line) onto a three-unit-cell structure with unit cells as in Fig. 7(a) calculated using FDFD. Here, the material loss in silver is included. In each unit cell the first stub is filled with silicon dioxide doped with CdSe quantum dots ($\varepsilon_1=4.0804-j0.0165$). The width and length of the first stub are $w_1=24$ nm and $h_1=90$ nm, respectively. The second and third stubs are both filled with InGaAsP with InAs quantum dots ($\varepsilon_2=11.38+j0.346$ and $\varepsilon_3=11.38+j0.838$). All other parameters are as in Fig. 2(b). (c) Amplitude of the transfer matrix elements $|D_{12}|$, $|D_{21}|$, and $|D_{22}|$ as a function of frequency for the three-unit-cell structure of Fig. 7(b).}
\end{figure}

To implement this in plasmonic waveguide-cavity systems, we consider a compound unit cell with three stubs which have the same resonance frequency $\omega_0$ and the same decay rate due to power escape through the waveguide $\frac{1}{\tau_0}$, but different decay (growth) rates due to the internal loss (gain) in the stub resonators $\frac{1}{\tau_i}, i=1, 2, 3$ [Fig. 7(a)]. We consider as before a finite structure consisting of three compound unit cells with $\phi=0$, $d=500$ nm, and $\eta=0.4$ [Eq. (1)]. Recall that the zeros of the reflection coefficient for the single-unit-cell structure $r_1$ are identical to the zeros of the reflection coefficient for the three-unit-cell structure $r_3$. We first use CMT to optimize the decay (growth) rates due to internal loss (gain) of all three stubs $\frac{1}{\tau_i}, i=1, 2, 3$, to simultaneously make the amplitude of the transfer matrix element $M_{21}$ as close to zero as possible, and the amplitude of the reflection coefficient in the right direction for the three-unit-cell structure $r_{3r}$ as large as possible. Using this approach, we find that $M_{21}$ vanishes and $r_{3r}$ diverges for $\frac{1}{\tau_1}=2.21\times10^{13} {\rm sec}^{-1}$, $\frac{1}{\tau_2}=1.81\times10^{11} {\rm sec}^{-1}$, and $\frac{1}{\tau_3}=-2.86\times10^{13} {\rm sec}^{-1}$. Thus, the first two stubs in the compound unit cell exhibit loss, while the third one exhibits gain. Fig. 7(b) shows the reflection spectra for the optimized three-unit-cell structure calculated using FDFD for incident waveguide modes from both the left and right directions. At $f=217$ THz the reflection from the left (black) is close to zero, while the reflection from the right (red) tends to infinity. In other words, the optimized structure simultaneously supports a unidirectional spectral singularity, as well as a TES on the right edge at $f=217$ THz.

In each compound unit cell, the first stub is filled with silicon dioxide doped with CdSe quantum dots ($\varepsilon_1=4.0804-j0.0165$). This active absorbing material is tunable, since the imaginary part of its refractive index can be modified with an external control beam \cite{Pacifici:07, Pacifici:09}. The width and length of the first stub are $w_1=24$ nm and $h_1=90$ nm, respectively. The second and third stubs are both filled with InGaAsP with InAs quantum dots but have dielectric constants with different imaginary parts ($\varepsilon_2=11.38+j0.346$ and $\varepsilon_3=11.38+j0.838$). The widths and lengths of these two stubs are $w_2=w_3=50$ nm and $h_2=h_3=60$ nm as before. As mentioned above, we choose the stub dimensions so that their resonance frequencies as well as their decay rates due to power escape through the waveguide are equal. In addition, taking into account the material loss of silver, we choose the imaginary parts of the dielectric constants of the materials filling the three stubs so as to satisfy the conditions $\frac{1}{\tau_1}=2.21\times10^{13} {\rm sec}^{-1}$, $\frac{1}{\tau_2}=1.81\times10^{11} {\rm sec}^{-1}$, and $\frac{1}{\tau_3}=-2.86\times10^{13} {\rm sec}^{-1}$, which, as mentioned above, were obtained using CMT.

If the overall transfer matrix of the optimized three-unit-cell structure is $\mathbf{D}=\left[
\begin{array}{c c}
D_{11} & D_{12} \\
D_{21} & D_{22} \\
\end{array}
\right]$, then the reflection coefficients for the left and right directions are given by $-\frac{D_{21}}{D_{22}}$ and $\frac{D_{12}}{D_{22}}$, respectively. A unidirectional reflectionless propagating mode for incidence from one side, as well as a unidirectional lasing mode for incidence from the other side is the key signature of a unidirectional spectral singularity. This exotic response has also been observed in $PT$-symmetric coupled cavity systems \cite{Ramezani:14}. In these systems, such a phenomenon is obtained for $D_{12} \rightarrow \infty$, $D_{21}\rightarrow 0$, and $D_{22} \neq 0$ at the operating frequency \cite{Ramezani:14}. Figure 7(c) shows the amplitudes of the transfer matrix elements $|D_{21}|$, $|D_{12}|$, and $|D_{22}|$ as a function of frequency for our optimized three-unit-cell structure. We observe that, in contrast to the $PT$-symmetric coupled cavity systems \cite{Ramezani:14}, the unidirectional spectral singularity in our case is obtained through
\begin{equation}
D_{12}\neq 0, \quad D_{21}\rightarrow 0, \quad D_{22}\rightarrow 0.
\end{equation}
$D_{12} \neq 0$ and $D_{22}\rightarrow 0$ [Fig. 7(c)] lead to diverging reflection from the right side  at $f=217$ THz [Fig. 7(b)]. $D_{21}$ approaching zero faster than $D_{22}$ [Fig. 7(c)] results
in vanishing reflection from the left side at $f=217$ THz [Fig. 7(b)].

\begin{figure}[htb]
\centering\includegraphics[width=10.5cm]{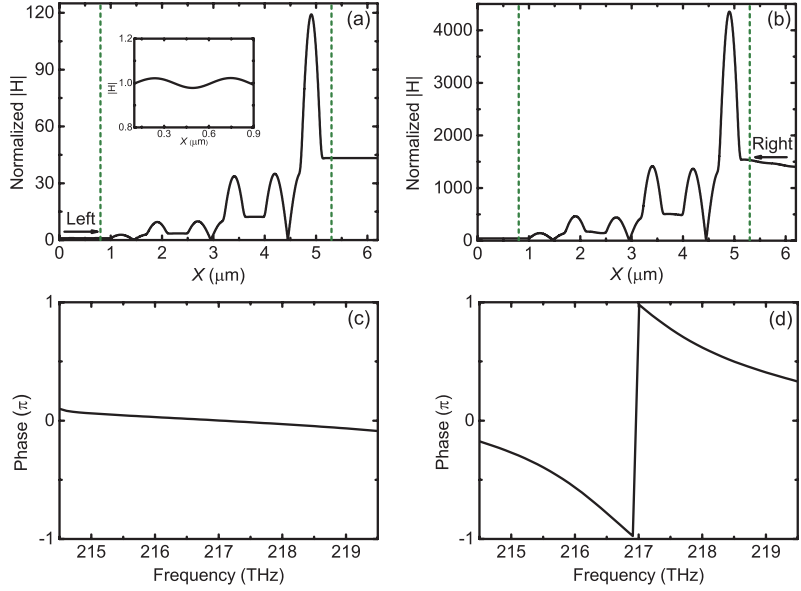}
\caption{(a) and (b) Profile of the magnetic field amplitude in the middle of the MDM waveguide, normalized with respect to the field amplitude of the incident waveguide mode in the middle of the waveguide, when the mode is incident from the left and right, respectively, onto the three-unit-cell structure of Fig. 7(b) at $f=217$ THz. The vertical dashed lines indicate the boundaries between the three-unit-cell structure and the MDM waveguide. The inset in Fig. 8(a) shows the normalized field profile for $0.1\mu$m $\leq X$ $\leq 0.9\mu$m, when the waveguide mode is incident from the left. (c) and (d) Spectra of the pPhase spectra of the  reflection coefficients when the mode is incident from the left and right, respectively, onto the three-unit-cell structure of Fig. 7(b) at $f=217$ THz. All other parameters are
as in Fig. 7(b).}
\end{figure}

In addition, we can observe both the TES and the unidirectional spectral singularity in the normalized magnetic field distributions for the optimized structure of Fig. 7(a) at $f=217$ THz [Figs. 8(a) and 8(b)]. When the waveguide mode is incident from the left, there is hardly any reflection, as seen in Fig. 8(a) and its inset. When the waveguide mode is incident from the right, the reflected wave is enhanced by three orders of magnitude [Fig. 8(b)]. In addition, the significantly enhanced field at the right edge of the structure verifies the existence of the edge state. In the previously reported $PT$-symmetric coupled cavity system \cite{Ramezani:14}, the main mechanism to create a unidirectional spectral singularity was based on Fano resonance trapping in the system. Light reflected from the gain side is strongly confined in the gain cavity and amplified, while light reflected from the lossy side is trapped in the lossy cavity and absorbed \cite{Ramezani:14}. In our case, the right lasing is caused by the topologically protected edge mode localization on the gain side (right side). However, due to the chirality of TESs in our system, there is no edge state on the lossy side (left side). The left reflectionlessness of our structure is originating from destructive interference rather than from light trapping. Thus, we observe that the field profiles in Figs. 8(a) and 8(b) are very similar except that the field enhancement is much larger for waveguide modes incident from the right. The strong light trapping results in a long light delay in the structure. In Figs. 8(c) and 8(d), we show the spectra of the phase of the reflection coefficient when the waveguide mode is incident from the left and right directions, respectively. In Fig. 8(d), we observe that the phase of the reflection coefficient for the optimized structure of Fig. 7(a) undergoes an abrupt jump at $f=217$ THz when the waveguide mode is incident from the right. The corresponding group delay experienced by the trapped light is given by $\tau_g=\frac{d\Phi(\lambda)}{d\lambda}$, where $\Phi$ is the phase of the reflection coefficient \cite{Ramezani:14,Huang:152}, and therefore diverges at $f=217$ THz. Thus, we confirm that the light reflected from the right side is strongly confined in the gain region and is amplified. In contrast, the phase of the reflection coefficient in the left direction does not undergo an abrupt jump, and varies smoothly with frequency [Fig. 8(c)]. It is worth noting that in this case the zero reflection in the left direction is not the signature of a TES on the left side, since the optimized structure includes a gain stub with $\frac{1}{\tau}<0$.

\begin{figure}[htb]
\centering\includegraphics[width=7cm]{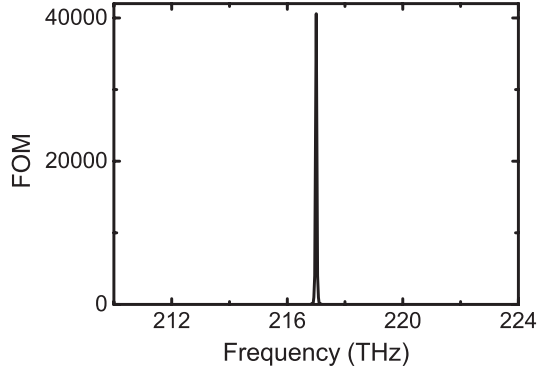}
\caption{FOM [Eq. (12)] for the optimized three-unit-cell structure with unit cells as in Fig. 7(a) as a function of frequency. All other parameters are as in Fig. 7(b).}
\end{figure}

The extremely sharp line shape of the reflection from the left side [Fig. 7(b)] provides an opportunity to design an extremely sensitive active structure at the unidirectional spectral singularity. To characterize the sensitivity of the optimized three-unit-cell structure with unit cells as in Fig. 7(a), we define the $figure$ $of$ $merit$ $(FOM)$ as the absolute value of the derivative of the reflection from the left $R_{3l}$ with respect to the imaginary part $\kappa$ of the refractive index of the active absorbing material filling the first stub of each unit cell [Fig. 7(a)] divided by $R_{3l}$
\begin{equation}
FOM=\Big|\frac{dR_{3l}}{R_{3l}d\kappa}\Big|,
\end{equation}
where $R_{3l}=|r_{3l}|^2$. The $FOM$ can be calculated using the following finite-difference approximation $\frac{dR_{3l}}{d\kappa}\simeq \frac{R_{3l}(\kappa+\Delta\kappa)-R_{3l}(\kappa-\Delta\kappa)}{2\Delta\kappa}$. In our calculations, we use $\Delta\kappa=10^{-4}<<\kappa$ \cite{Huang:192}. Figure 8 shows the calculated $FOM$ as a function of frequency. The maximum value of the $FOM$ is $\sim 40500$ at the unidirectional spectral singularity ($f=217$ THz), which is two orders of magnitude larger than the $FOM$ in plasmonic sensors based on the Fano resonance \cite{Lu:12}.

We note that our choice for the imaginary part of the refractive index of the active absorbing material ($\kappa=0.0165$) is within the range of experimentally achievable values \cite{Pacifici:07, Pacifici:09, Min:09}. In addition, the imaginary parts of the refractive index of the gain material filling the second and third stubs of each unit cell in the optimized structure [Fig. 7(a)] are 0.346 and 0.838, respectively, which correspond to gain coefficients of $g\approx 4666$ cm$^{-1}$ and $g\approx 11250$ cm$^{-1}$, respectively \cite{Nezhad:04}. These can be realized with ultra-high-density quantum dot structures \cite{Bimberg:98, Akahane:11}.

\section{Conclusions}
In this paper, we designed non-Hermitian periodic plasmonic waveguide-cavity structures based on the AAH model to realize both a TES and a singular point at the same frequency. We used the transfer matrix method and CMT to account for the behavior of the proposed structures. We first showed that we can realize both a TES and an EP at the same frequency when a proper amount of loss is introduced into the plasmonic structure. We also showed that the structure can exhibit both a TES and a SS at the same frequency when a proper amount of gain is introduced into the structure. In addition, we showed that we can realize both unidirectional spectral singularities and TESs when a proper amount of loss and gain are introduced into the plasmonic structure. The optimized structure supports unidirectional reflectionless propagation for incidence from one side, as well as a unidirectional lasing for incidence from the other side. The underlying physical mechanism of the unidirectional lasing is the topologically protected edge mode localization on the gain side, while the unidirectional reflectionlessness is originating from destructive interference. Finally, we found that for such a structure the sensitivity of the reflection to variations of the refractive index of the active material, when the waveguide mode is incident from the side which supports unidirectional reflectionlessness, is significantly enhanced at the unidirectional spectral singularity. Thus, the optimized structure operating at the unidirectional spectral singularity can lead to extremely sensitive active photonic devices such as modulators and switches.

As final remarks, our results demonstrate the connection between topologically protected edge states and different types of singular points, and could potentially contribute to the development of a new generation of singularity-based plasmonic devices with enhanced performance. The concept of combining gain and loss to realize both TESs and singular points at the same frequency could also be applied in other photonic and acoustic systems. In addition, we note that TESs at singular points could be realized in three-dimensional plasmonic waveguide-cavity systems based on plasmonic coaxial waveguides \cite{Shin:13, Mahigir:15}.

\section*{Funding}
National Natural Science Foundation of China (61605252); National Key Research and Development Program of China (2019YFA0706301); National Natural Science Foundation of China (12004446).

\section*{Disclosures}
The authors declare no conflicts of interest.


%
%

\end{document}